\documentstyle[12pt]{article}

\begin{document}

\begin{flushright}
IMSc/2000/10/55 \\ 
hep-th/0010121 
\end{flushright} 

\vspace{2mm}

\vspace{2ex}

\begin{center}
{\large \bf Brane World Scenario with $m$-form field: } \\ 

\vspace{2ex}

{\large \bf Stabilisation of Radion Modulus and Self Tuning} \\

\vspace{2ex}

{\large \bf Solutions } \\

\vspace{8ex}

{\large  S. Kalyana Rama}

\vspace{3ex}

Institute of Mathematical Sciences, C. I. T. Campus, 

Taramani, CHENNAI 600 113, India. 

\vspace{1ex}

email: krama@imsc.ernet.in \\ 

\end{center}

\vspace{6ex}

\centerline{ABSTRACT}
\begin{quote} 
We consider (n + m + 1) dimensional bulk spacetime,
containing flat (n + m - 1) dimensional parallel branes, with
topology ${\bf R^{n - 1} \times T^m}$. We assume that the
graviton and an $m$-form field are the only bulk fields and that
the $m$-form field has non vanishing components along the ${\bf
T^m}$ directions only.  We then find that the $m$-form field,
with suitable bulk and brane potentials, can stabilise the
radion modulus at the required value with no fine tuning.  We
find self tuning solutions also.
\end{quote}

\vspace{2ex}


\newpage

\vspace{4ex}

{\bf 1.}  
Randall and Sundrum have recently proposed a simple brane
world scenario to solve the hierarchy problem \cite{rs1}. The
bulk spacetime in this scenario is five dimensional, with the
extra spatial dimension having the topology ${\bf S^1/Z_2}$.
There are two $3$-branes located at the fixed points of the
${\bf Z_2}$. One of them, called the visible brane, is assumed
to represent our universe.  The standard model fields are
localised on the visible brane, whereas the graviton 
is a bulk field.  For a suitable choice of constant 
bulk and brane potentials, Randall and Sundrum 
find a warped solution where the brane metric
varies exponentially. This introduces an exponentially large
difference in the scales on the branes which, essentially,
solves the hierarchy problem. See \cite{rs1} for details.

The exponential difference in the scales depends crucially on
the radion modulus which corresponds to the size of ${\bf S^1}$
and which needs to be stabilised at the required value, 
$\simeq 40 \; \times$ {\em the fundamental scale}, with no fine
tuning.  Soon after the work of Randall and Sundrum, 
Goldberger and Wise have
indeed proposed an elegant mechanism for such stabilisation
\cite{gw}.  In this model, an extra bulk scalar field, with
suitable bulk and brane potentials, stabilises the modulus with
no fine tuning of the parameters involved. For details, see
\cite{gw,gubser}. See also \cite{hierarchy} for other solutions
of the hierarchy problem with two extra dimensions.

Of course, fundamental theories, such as string theory or
supergravity theories from which such models may perhaps be
derived \cite{derive}, do contain extra dimensions and many bulk
scalar fields. However, they also contain numerous $m$-form
antisymmetric bulk fields, $m \ge 1$. It is therefore worthwhile
to explore the consequences of such an $m$-form bulk
field\footnote{The $m$-form fields have been considered in
recent studies \cite{pform} in other contexts.} and study,
for example, whether such a field can also stabilise the radion
modulus at the required value with no fine tuning.

In this letter, we study the brane world scenario with an
$m$-form antisymmetric bulk field, $m \ge 1$.  We consider
$D = (n + m + 1)$ dimensional bulk spacetime, containing flat
$(n + m - 1)$ dimensional parallel branes, with topology 
${\bf R^{n - 1} \times T^m}$. In the present work, we assume
that the graviton and an $m$-form field are the only bulk fields
and that the $m$-form field has non vanishing components along
the ${\bf T^m}$ directions only. We then study 
some of its consequences.

The equations of motion that follow are difficult to solve
analytically. Nevertheless, it turns out that the qualitative
features of the solutions can be understood in certain 
cases. We find, by an analysis similar to that of \cite{gw},
that the $m$-form field, with suitable bulk and brane
potentials, can indeed stabilise the radion modulus at the
required value, with no fine tuning of the parameters involved.

Also, self tuning solutions have been found recently in the
presence of bulk scalars \cite{kachru}. In these solutions, the
brane world admits flat metric even when the brane potential 
is non zero, but at the cost of a singularity in the extra
dimension \cite{kachru}.  Figuratively speaking, the brane
metric tunes itself to be flat for different values of the brane
potential by moving the singularities around.

We find such self tuning solutions also in the presence of an
$m$-form antisymmetric bulk field, $m \ge 1$.  However, these
solutions differ in details from those in \cite{kachru}.

The plan of the paper is as follows. We first present our set
up. We then show that an $m$-form bulk field can indeed
stabilise the radion modulus. We then present the self tuning
solutions, and conclude with a few remarks.

{\bf 2.}  
We consider $D = (n + m + 1)$ dimensional spacetime, $m \ge 1$,
containing flat $(n + m - 1)$ dimensional branes, with topology
${\bf R^{n - 1} \times T^m}$. We assume that ${\bf T^m}$ is of
$D$ dimensional planckian size, and that one of the branes
represents our universe.  Thus, on a macroscopic scale, the
branes are $(n - 1)$ dimensional, with $n = 4$ corresponding 
to the observable case. The bulk fields are taken to be 
the metric $g_{M N}$ and a totally antisymmetric $m$-form 
field $B_{M_1, \cdots, M_m}$. 
The relevent action can be written as 
\begin{equation}\label{s+sbr}
S = \int d^D x \sqrt{- g} \left( \frac{R}{4} 
- \frac{G^2}{2 (m + 1)!} - V(\chi) 
- \sum_I \delta(y - y_I) \Lambda(\chi) \right) \; , 
\end{equation} 
where $g = det(g_{MN})$, $G = d B$ is the $(m + 1)$-form field
strength for $B_{M_1, \cdots, M_m}$, $V$ and $\Lambda_I$ are 
the brane and bulk potentials respectively which are, in general, 
functions of the diffeomorphism invariant quantity 
$\chi \equiv B_{M_1, \cdots M_m} B^{M_1, \cdots M_m}$, and 
$y_I$ are the locations of the branes. 
In our notation, $x^M = (x^\mu, \xi^i, y)$, 
$\mu = 0, 1, \cdots, (n - 1)$, and $i = 1, \cdots, m$ denote the
$D$ dimensional spacetime coordinates, $(x^\mu, \xi^i)$ the
brane worldvolume coordinates, and $y$ the transverse spatial
coordinate. The signature of the metric is $(-, +, +, \cdots)$.
The Riemann tensor is $R^M_{NKL} = \partial_K \Gamma^M_{NL} 
+ \cdots$. The $D$ dimensional planck mass is set to unity, and
the dimensionful quantities, here and in the following, are all
taken to be of ${\cal O}(1)$ unless mentioned otherwise. 

We look for warped solutions, with the $m$-form field having
components only along the ${\bf T^m}$ directions. Explicitly, 
our ansatz is given by 
\begin{eqnarray}
& & d s^2 = g_{M N} d x^M dx^N = e^{2 A(y)} \eta_{\mu \nu} 
d x^\mu d x^\nu + e^{2 B(y)} \delta_{i j} 
d \xi^i d \xi^j + d y^2 
\nonumber \\ 
& & B_{n,(n + 1), \cdots, (n + m - 1)} = \beta(y) \; , 
\label{ansatz} 
\end{eqnarray} 
where $\eta_{\mu \nu} = diag(- 1, 1, 1, \cdots)$ and all other
independent components of the $m$-form field $B_{M_1, \cdots,
M_m}$ are set to zero. Then, $\chi = m! e^{- 2 m B} \beta^2$. 
The bulk equations of motion\footnote{Note that equations
(\ref{beta})-(\ref{energy}) are not all independent.
Differentiating equation (\ref{energy}) yields an identity, upon
using equations (\ref{beta})-(\ref{sigma}).} are given by 
\begin{eqnarray}
& & \beta_{yy} + (\Sigma_y - 2 m B_y) \beta_y 
- 2 m! V_\chi \beta = 0 \label{beta} \\
& & A_{yy} + A_y \Sigma_y = \frac{2}{D - 2} \; 
(m e^{- 2 m B} \beta_y^2 
+ 2 m \chi V_\chi - 2 V)  \label{a} \\
& & \Sigma_{yy} + \Sigma_y^2 = \frac{2}{D - 2} \; 
(m e^{- 2 m B} \beta_y^2 
+ 2 m \chi V_\chi - 2 (n + m) V)  \label{sigma} \\
& & \Sigma_y^2 - n A_y^2 - m B_y^2  = 
2 e^{- 2 m B} \beta_y^2 - 4 V \label{energy} \; , 
\end{eqnarray}
where $( \; )_y \equiv \frac{\partial}{\partial y}( \; )$, 
$( \; )_\chi \equiv \frac{\partial}{\partial \chi}( \; )$, 
and $\Sigma \equiv ln \sqrt{- g} = n A + m B$. The independent
fields are thus $\beta$, $A$, and $\Sigma$, or equivalently $B$.
The branes introduce $\delta$-function source terms in equations
(\ref{beta})-(\ref{sigma}), but not in (\ref{energy}), which in
turn introduce discontinuities in various derivatives at the
brane locations. For example, a brane with potential $\Lambda$
and located at $y = 0$ introduces the following discontinuities:
\begin{eqnarray}
{[} \beta_y {]} & = & 2 m! \Lambda_\chi \beta \nonumber \\
{[} A_y {]} & = & \frac{2}{D - 2} \; 
(2 m \chi \Lambda_\chi - \Lambda) \nonumber \\ 
{[} \Sigma_y {]} & = & \frac{2}{D - 2} \; 
(2 m \chi \Lambda_\chi - (n + m) \Lambda) \; , \label{disc}
\end{eqnarray} 
where ${[} X {]} = X(0+) - X(0-)$ denotes the discontinuity in
the quantity 
$X$ and the quantities on the right hand side are evaluated at
$y = 0$. Note that if $\Lambda = constant$ then 
$\Lambda_\chi = 0$ and, hence, ${[} \beta_y {]} = 0$ and 
${[} A_y {]} = \frac{{[} \Sigma_y {]}}{n + m} 
= - \frac{2 \Lambda}{D - 2}$. Evaluating equation 
(\ref{energy}) at $y = 0+$ and $y = 0-$ then gives 
\begin{equation}\label{discsigma}
\Sigma_y(0+) = - \Sigma_y(0-) = 
- \frac{(n + m) \Lambda}{D - 2} \; . 
\end{equation} 

In this letter, we study the following two cases: \\
{\bf (i)} 
The transverse direction is a circle, with 
$- y_1 \le y \le y_1$. The points $(x^\mu, \xi^i, y)$ and
$(x^\mu, \xi^i, - y)$ are identified. Two branes, with brane
potentials $\Lambda_0(\chi)$ and $\Lambda_1(\chi)$, are located
respectively at $y = 0$ and $y = y_1$, the fixed points of 
the above identification. This is the case considered in 
\cite{rs1}. \\ 
{\bf (ii)} 
The transverse direction is infinite, $- \infty \le y \le
\infty$. A single brane is located at $y = 0$, with brane
potential $\Lambda(\chi)$. This is the case considered in 
\cite{kachru}. 

In the following, we set $A = B = \Sigma = 0$ at $y = 0$ with no
loss of generality. Also, we analyse the solutions for $y > 0$
only.  The analysis for $y < 0$ follows similarly, with the
initial conditions at $y = 0-$ fixed by those at $y = 0+$ and
the discontinuity equations (\ref{disc}).

{\bf 3.}  
Consider now the case {\bf (i)}. Let $\beta = 0$ and
$V = - V_0$ where $V_0$ is a positive constant. Then the
warped bulk solution \cite{rs1} is given by 
\begin{equation}\label{rs}
A = \frac{\Sigma}{n + m} = - k |y| 
\end{equation}
where the constant $k$, taken to be positive, is given by 
\[
k^2 = \frac{4 V_0}{(D - 1)(D - 2)} \; . 
\]
Since the points $(x^\mu, \xi^i, y)$ and $(x^\mu, \xi^i, - y)$
are identified, the discontinuities (\ref{disc}) at $y = 0$ and
$y = y_1$ then imply that 
\[
\Lambda_0 = - \Lambda_1 = (D - 2) k \; . 
\]
These conditions amount to two fine tunings, one equivalent to
the stabilisation of the radion modulus $y_1$, and another
equivalent to the vanishing of the cosmological constant
\cite{gw,gubser}.  From equation (\ref{rs}), it follows that the
ratio of the scales on the (visible) brane at $y = y_1$ to
the scales on the brane at $y = 0$ is given by 
$e^{- k y_1}$. Thus, if $k y_1 \simeq 40$ then $e^{- k y_1} 
\simeq 10^{- 16}$.  Essentially, it is this exponential
difference in the scales which solves the hierarchy problem
\cite{rs1}.

However, to really solve the hierarchy problem, the modulus
$y_1$ needs to be stabilised at the required value with no fine
tuning. To this end, Goldberger and Wise have proposed an
elegant mechanism where a bulk scalar field, with suitable bulk
and brane potentials, stabilises the modulus $y_1$. The required
value of $y_1$ can then be achieved with no fine tuning
\cite{gw}. See \cite{gubser} also for a complete analysis.

Following \cite{gw}, we study here whether the $m$-form field,
with suitable bulk and brane potentials $V(\chi)$,
$\Lambda_0(\chi)$ and $\Lambda_1(\chi)$, can also stabilise 
the modulus $y_1$. Therefore, now, let $\beta \ne 0$, and 
$V(\chi) \ne constant$. Equations (\ref{beta})-(\ref{energy})
are difficult to solve analytically for any non trivial
potential\footnote{ The solution generating technique of
\cite{gubser}, used for the bulk scalar, is not applicable
here.}.  Nevertheless, following the analysis of \cite{gw}, it
is possible to study the stabilisation of $y_1$ by the $m$-form
field.

Let the bulk potential be given by 
$V(\chi) = - V_0 + \frac{\Omega^2 \chi}{2 m!}$, where $V_0$ 
and $\Omega^2$ are positive constants.  We assume that $V_0$
dominates the second term, and ensure later that the solutions
obtained are consistent with this assumption. The background
fields $A(y)$ and $\Sigma(y)$, equivalently $B(y)$, are then
determined by $V_0$ and are given in (\ref{rs}). The brane
potentials $\Lambda_0(\chi)$ and $\Lambda_1(\chi)$ are taken 
to be such as to enforce the boundary conditions\footnote{For
example, $\Lambda_I = \lambda_I (\chi^2 - \chi_I^2)^2$, 
$I = 0, 1$, with $\lambda_I$ constant and very large. The
effects of finite $\lambda_I$ can be analysed as in \cite{gw},
and they do not change the results significantly.} 
\begin{eqnarray*}
& & \chi_0 \equiv \chi(0) = m! \beta_0^2 \\
& & \chi_1 \equiv \chi(y_1) = m! 
e^{2 m k y_1} \beta_1^2  \; , 
\end{eqnarray*}
with $\chi_0 \ne \chi_1$.  
It then follows that $\beta(y)$ satisfies the equation 
\[
\beta_{yy} - (n - m) k \beta_y - \Omega^2 \beta = 0 
\]
with the boundary conditions $\beta(0) = \beta_0$ and
$\beta(y_1) = \beta_1$. Hence, $\beta(y)$ is given by
\begin{equation}\label{betagw}
\beta = \left( \frac{\beta_1 - \beta_0 e^{k_2 y_1}}
{1- e^{(k_2 - k_1)y_1}} \right) \; e^{k_1 (y - y_1)}
\; + \; \left( \frac{\beta_0 - \beta_1 e^{- k_1 y_1}} 
{1- e^{(k_2 - k_1)y_1}} \right) \; e^{k_2 y} \; , 
\end{equation}
where $k_1$ and $k_2$ ($< 0$) are given by 
\[
\frac{k_1}{k} = \frac{n - m}{2} + \nu \; , \; \; \; 
\frac{k_2}{k} = \frac{n - m}{2} - \nu \; , \; \; \; 
\nu \equiv \sqrt{\frac{(n - m)^2}{4} 
+ \frac{\Omega^2}{k^2}} \; . 
\]

We now check whether the back reaction of $\beta$ on $A$ and
$\Sigma$ can be consistently neglected. The energy momentum
tensor $T_M^N (\beta)$ for $\beta$ is given by (no summation
over repeated indices) 
\[
T^y_y (\beta) = e^{- 2 m B} (\beta_y^2 - \Omega^2 \beta^2)  
\; , \; \; \; T^i_i (\beta) = - T^\mu_\mu (\beta) = 
e^{- 2 m B} (\beta_y^2 + \Omega^2 \beta^2) \; . 
\]
Substituting the solution for $\beta$, and writing $T_M^N
(\beta)$ in terms of the diffeomorphism invariant quantity
$\chi$, it can be immediately seen that the energy momentum
tensor $T_M^N (\beta)$ of the $m$-form field is of the order
$k^2 \chi_{0(1)}$. Thus, if $\chi_0, \chi_1 \ll 1$, in units
where the $D$ dimensional planck mass is set to unity, then
$|T_M^N (\beta)| \ll V_0$ and the back reaction of $\beta$ on
$A$ and $\Sigma$ can be consistently neglected. Hence, in the
following, we assume that $\chi_0, \chi_1 \ll 1$, ensuring 
thus that the back reaction can be consistently neglected. 

Substituting the solution for the $m$-form field into the bulk
action and integrating over the $y$-coordinate then yields an
effective potential, $V_{eff}(y_1)$, for the modulus $y_1$.
In terms of $\chi$, it is given by 
\begin{equation}\label{veff}
V_{eff}(y_1) = \frac{v_1(y_1) + v_2(y_1)}
{m ! (1 - e^{- 2 \nu k y_1})} \; , 
\end{equation}
where the terms $v_1(y_1)$ and $v_2(y_1)$, both non 
negative, are given by 
\begin{eqnarray*}
& & v_1(y_1) = k_1 (\sqrt{\chi}_1 - \sqrt{\chi}_0 
e^{K_2 y_1})^2 \; e^{- (n + m) k y_1} \\
& & v_2(y_1) =  - k_2 (\sqrt{\chi}_0 - \sqrt{\chi}_1 
e^{- K_1 y_1})^2 \; , 
\end{eqnarray*} 
where $K_1 = k_1 + m k$ and $K_2 = k_2 + m k$. Note that 
$K_1 \ge n k > 0$ and $K_1 K_2 = (n m k^2 - \Omega^2)$. 
Hence, $K_2$ has the same sign as $(n m k^2 - \Omega^2)$.

Although $V_{eff}(y_1)$ is a complicated function of $y_1$, its
qualitative behaviour can be easily understood. Note that
$V_{eff}(y_1) > 0$ always. Also, 
$V_{eff}(0) \to \infty$, and 
$V_{eff}(\infty) \to - \frac{k_2 \chi_0}{m!}$ from above. 
Therefore, if $V_{eff}(y_1) < V_{eff}(\infty)$ for some 
positive $y_1$ then it follows that $V_{eff}$ must have 
a minimum. Indeed, let 
\begin{equation}\label{y1c}
y_{1c} = \frac{1}{2 K_2} \; ln \frac{\chi_1}{\chi_0} \; . 
\end{equation}
For the sake of definiteness, let us choose in the following 
$\chi_0 > \chi_1$ 
and $K_2 < 0$, equivalently $\Omega^2 > n m k^2$. 
Then $y_{1c}$ is positive, $v_1(y_{1c}) = 0$, and 
\[
V_{eff}(y_{1c}) = (1 - e^{- 2 \nu k y_{1c}}) \; 
V_{eff}(\infty) \; \; < V_{eff}(\infty) \; . 
\]
Since $y_{1c} > 0$ for the above choice of parameters 
and $V_{eff}(y_{1c}) < V_{eff}(\infty)$, it follows that 
$V_{eff}$ must have a minimum. Also, 
\[
\frac{d V_{eff}}{d y_1} (y_{1c}) = - \frac{(n + m)}{m!} \; 
k k_2 \sqrt{\chi_0 \chi_1} e^{- K_1 y_{1c}} \; , 
\] 
which is small since $\chi_0, \chi_1 \ll 1$ by assumption and,
moreover, $K_1 y_{1c} > 0$.  Therefore, to a good approximation,
one may take $y_{1c}$ itself to be the minimum of $V_{eff}$ and,
hence, to be the stabilised value of the modulus $y_1$. This
approximation improves vastly for the case of our interest, as
we will see now.

As can be seen from equation (\ref{y1c}), a large value of 
$k y_{1c}$ can be easily achieved by choosing $\frac{K_2}{k}$
to be sufficiently small. For example, the choice 
\[
\chi_0 = e^2 \chi_1 \simeq 10 \chi_1 
\; , \; \; \; 
\Omega^2 \simeq (n m + \frac{n + m}{40}) k^2 \; , 
\]
gives $k y_{1c} \simeq 40$, a phenomenologically interesting
value which solves the hierarchy problem.  The above choice
amounts to a tuning upto only a couple of orders of magnitude
and, hence, is quite acceptable. Also, note that 
$K_1 y_{1c} > n k y_{1c} \gg 1$ and, hence, 
$\frac{d V_{eff}}{d y_1} (y_{1c}) \simeq 0$. This shows that, to
an excellent approximation, one may indeed take $y_{1c}$ itself
to be the minimum of $V_{eff}$ and, hence, to be the stabilised
value of the modulus $y_1$.

Thus, it follows that the $m$-form field, with suitable 
bulk and brane potentials $V(\chi)$, $\Lambda_0(\chi)$ and
$\Lambda_1(\chi)$, can also stabilise the modulus $y_1$ 
by a mechanism similar to that of \cite{gw}. The required 
value of the stabilised modulus can be easily achieved 
with no fine tuning.

{\bf 4.}
Consider now the case {\bf (ii)} where the transverse direction
is infinite, $- \infty \le y \le \infty$, and a single brane is
located at $y = 0$, with brane potential $\Lambda(\chi)$. Let
$\beta \ne 0$ and the bulk potential $V = 0$. We now look
for self tuning solutions of the type found in \cite{kachru}.

Equation (\ref{beta}) can be solved to give 
\begin{equation}\label{betak}
\beta_y = \sqrt{\frac{D - 2}{2}} \; K e^{m B - n A} 
\; \; \; {\rm and,} \; \; {\rm hence,} \; \; \; 
\frac{2 e^{- 2 m B} \beta_y^2}{D - 2} = K^2 e^{- 2 n A} 
\end{equation} 
where $K$ is an arbitrary constant. We now need to solve
equations (\ref{a})-(\ref{energy}). Let us define 
\begin{equation}\label{f}
F(y) = \frac{A_y}{\Sigma_y - A_y} 
\; \; \; \longleftrightarrow \; \; \; 
\Sigma_y = \frac{(1 + F) A_y}{F} \; . 
\end{equation}
Thus, if $A$ and $F$ are known then $\Sigma$, equivalently $B$,
and $\beta$ can be determined by direct integrations. Let the
initial values of $A_y, \Sigma_y, \beta_y$, and $F$ at $y = 0+$
be $A_y(0), \Sigma_y(0), \beta_y(0)$, and $F_0$ respectively.
Equation (\ref{f}) relates $A_y(0), \Sigma_y(0)$ and $F_0$, and
equation (\ref{energy}) determines $\beta_y(0)$, equivalently
$K$, upto a sign.  The discontinuity equations (\ref{disc}) will
then determine the initial values of the fields at $y = 0-$. 

From equations (\ref{a}), (\ref{sigma}), (\ref{energy}), 
and (\ref{f}) one obtains that 
\begin{eqnarray}
& & F_y A_y =  m K^2 e^{- 2 n A} \; F \\ 
& & A_y^2 = \frac{m K^2 e^{- 2 n A} F^2}
{(n - 1) (F - F_1) (F_2 - F)}  \; , \label{ayf}
\end{eqnarray}
where $F_1$ and $F_2$ are given by 
\[
F_1 = \frac{1 - \sqrt{x}}{n - 1} \; , \; \; \; 
F_2 = \frac{1 + \sqrt{x}}{n - 1} \; , \; \; \; 
x \equiv \frac{n m }{n + m - 1} \; . 
\]
Note that $- 1 < F_1 \le 0$ and $F_2 > 0$. Also, 
$F(y) \in {[}F_1, F_2{]}$ since $K^2 \ge 0$ in equation
(\ref{ayf}). We assume that $A_y(0) \ne 0$ and  
$F_0 \in (F_1, F_2)$, so that $K \ne 0$ in the following. From
the above equations, it follows that 
\begin{equation}\label{fyay}
\frac{d A}{d F} = \frac{A_y}{F_y} 
= \frac{F}{(n - 1) (F - F_1) (F_2 - F)}
\end{equation}
which, together with the initial conditions described 
before, can be integrated to give 
\begin{equation}\label{af}
A = \frac{F_1}{2 \sqrt{x}} ln \frac{F - F_1}{F_0 - F_1} 
- \frac{F_2}{2 \sqrt{x}} ln \frac{F_2 - F}{F_2 - F_0} \; . 
\end{equation}

Equation (\ref{af}) describes the behaviour of $A$ as a function
of $F$: $A(F_2) \to \infty$. If $m > 1$ then $x > 1$ and 
$F_1 < 0$. Then, $A(F_1) \to \infty$ and $A(F)$ has a minimum 
at $F = 0$, given by 
\[
A_{min} = \frac{F_1}{2 \sqrt{x}} ln \frac{- F_1}{F_0 - F_1} 
- \frac{F_2}{2 \sqrt{x}} ln \frac{F_2}{F_2 - F_0} \; . 
\] 
If $m = 1$ then $x = 1$ and $F_1 = 0$. Then, 
$A(F_1) \to A_{min}$, given above but now with $F_1 = 0$.

From equations (\ref{ayf})-(\ref{af}) we obtain,  
after a straightforward algebra, 
\begin{equation}\label{fy} 
F_y = F_y(0) \; 
\left( \frac{F - F_1}{F_0 - F_1} \right)^{f_1} \; 
\left( \frac{F_2 - F}{F_2 - F_0} \right)^{f_2} 
\end{equation} 
where $F_y(0)$ is the value of $F_y$ at $y = 0+$, and 
\[
f_1 = \frac{(2 n - 1) \sqrt{x} - n}{2 (n - 1) \sqrt{x}} 
\; \; < 1 \; , \; \; \; 
f_2 = \frac{(2 n - 1) \sqrt{x} + n}{2 (n - 1) \sqrt{x}} 
\; \; > 1 \; . 
\] 
Thus, $F(y)$ and, hence, $A(y)$ can in principle be obtained as
explicit functions of $y$. But explicit expressions are
cumbersome and, also, are not illuminating. Nevertheless, the
qualitative features of the solutions can all be obtained from
the above equations only, as follows.

From equations (\ref{fy}), (\ref{fyay}), (\ref{f}), and 
the fact that $- 1 < F_1 < F_0, \; F < F_2$ and, hence, 
$(1 + F_0) > 0$, we have 
\begin{equation}\label{sgn}
sgn F_y = sgn F_y(0) 
= sgn \frac{A_y(0)}{F_0} 
= sgn \frac{\Sigma_y(0)}{1 + F_0} = sgn \Sigma_y(0) \; . 
\end{equation} 
Also, for a given value of $F_0$, the effect of $|\Sigma_y(0)|$
is simply to scale the variable $y$ in $A(y)$ and $\Sigma(y)$,
see equation (\ref{fy}). It is this feature that leads to 
self tuning. 

Thus, if $\Sigma_y(0) > 0$ then $F_y > 0$ and, as $y$ increases,
$F \to F_2$, and $A \to \infty$. If $\Sigma_y(0) < 0$ then 
$F_y < 0$ and, as $y$ increases, $F \to F_1$, and $A \to \infty$
(or, $A_{min}$) when $m > 1$ (or, $m = 1$). 
For example, let $F_0 > 0$. If $A_y(0) > 0$ then 
$\Sigma_y(0) > 0$ since $1 + F_0 > 0$.  Then, $F \to F_2$, 
and $A$ increases to $\infty$. If $A_y(0) < 0$ then 
$\Sigma_y(0) < 0$ since $1 + F_0 > 0$.  Then, $F \to F_1$, and
$A$ decreases, reaches a minimum and, for $m > 1$, increases
again to $\infty$. Furthermore, for a given $F_0$, the value of
$y \equiv y(F_{1(2)})$ at which $F = F_{1(2)}$ simply scales
with $|\Sigma_y(0)|$.

Consider now the solutions as $F \to F_{1(2)}$. Firstly, it 
can be shown, by a straightforward algebra, that in this limit
$\beta$ approaches a finite constant, $\beta_{1(2)}$, and that 
$\beta - \beta_{1(2)} \propto |F - F_{1(2)}|$. Now consider
equation (\ref{fy}). In the limit $F \to F_2$, it can be 
written as 
\[
F_y \simeq {\cal F}_2 (F_2 - F)^{f_2} 
\]
for some positive constant ${\cal F}_2$. 
The solution is then given by 
\[
(F_2 - F)^{1 - f_2} = k y + constant 
\]
where $k \equiv (f_2 - 1) {\cal F}_2$. Since $f_2 > 1$, 
it follows that $k > 0$, and that $y \to \infty$ as 
$F \to F_2$. Hence, $y(F_2) = \infty$. 
Equations (\ref{af}) and (\ref{f}) then give 
\begin{eqnarray} 
A & = & constant + \frac{1 + \sqrt{x}}{n + \sqrt{x}}
\; ln \; y \nonumber \\
B & = & constant - \frac{(n - 1) \sqrt{x}}{m (n + \sqrt{x})}
\; ln \; y \; .  \label{nosing} 
\end{eqnarray}
Thus, as $F \to F_2$, we have that $y \to y(F_2) = \infty$, 
$A \to \infty$, and $B \to - \infty$.  Also, as can be easily
checked, the curvature invariants are all finite and non
singular in this limit.

Let $m > 1$. In the limit $F \to F_1$, equation (\ref{fy}) 
can be written as 
\[
F_y \simeq - {\cal F}_1 (F - F_1)^{f_1} 
\]
for some positive constant ${\cal F}_1$. 
The solution is then given by 
\[
(F - F_1)^{1 - f_1} = k (y - y(F_1)) 
\]
where $k \equiv (f_1 - 1) {\cal F}_1$. Since $f_1 < 1$, it
follows that $k < 0$. Therefore, as $F \to F_1$, $y \to y(F_1)$,
a finite positive value. Equations (\ref{af}) and (\ref{f}) then
give
\begin{eqnarray}
A & = & constant - \frac{\sqrt{x} - 1}{n - \sqrt{x}}
\; ln (y(F_1) - y) \nonumber \\ 
B & = & constant + \frac{(n - 1) \sqrt{x}}{m (n - \sqrt{x})}
\; ln (y(F_1) - y) \; .  \label{sing}   
\end{eqnarray} 
Thus, as $F \to F_1$, we have that $y \to y(F_1)$, 
$A \to \infty$, and $B \to - \infty$. Also, as can be 
easily checked, the Ricci scalar $R$ remains finite, but 
$R_{MNPQ} R^{MNPQ}$ diverges in this limit. Thus, there is a
singularity at $y(F_1)$.  One may then assume that the $y$-axis
is cut off at this point.  However, see \cite{nilles} for 
detailed discussions and interpretations of these singularities.

If $m = 1$ then $x = 1$, $F_1 = 0$, and $f_1 = \frac{1}{2}$. 
Repeating the analysis given above, one finds that, in the limit
$F \to F_1 = 0$, $F(y)$ and $B(y)$ are given as in the $m > 1$
case above, whereas $A(y)$ is now given by 
\[
A - A_{min} \propto k^2 (y(F_1) - y)^2 \; . 
\]
Thus, as $F \to F_1$, we have that $y \to y(F_1)$, 
$A \to A_{min}$, and $B \to - \infty$. The line 
element (\ref{ansatz}) is given, in this limit, by 
\[
d s^2 = (constant) \eta_{\mu \nu} d x^\mu d x^\nu 
+ k^2 (y(F_1) - y)^2 d \xi^2 + d y^2 \; ,  
\]
from which it follows that there is generically a conical
singularity at $y(F_1)$ since generically $k \xi \ne 2 \pi$.
One may then assume that the $y$-axis is cut off at this
point. See \cite{hierarchy} where also similar conical
singularities arise.

The $n$ dimensional effective planck mass, $M_n$, is given by  
\[
M_n^{n - 2} = Vol({\bf T^m}) \; 
\int d y \; e^{\Sigma - 2 A} \; , 
\]
where $Vol({\bf T^m})$ is the volume of ${\bf T^m}$, and is of 
${\cal O}(1)$ in our units. From the expressions given above, it
follows that $M_n$ diverges in the limit $F \to F_2$, whereas it
approaches a constant in the limit $F \to F_1$.

To see clearly the self tuning features of these solutions, 
let $\Lambda(\chi) = constant$. Then $\Sigma_y(0)$
is related to $\Lambda$ by equation (\ref{discsigma}).
Then, as described below equation (\ref{sgn}), the effect of
$|\Lambda|$ is simply to scale the $y$ coordinate. Thus, if
$\Lambda > 0$ then $\Sigma_y(0) < 0$ and, hence, $F_y < 0$. 
Then, $F \to F_1$ and the $n$ dimensional Planck mass is finite,
but there is a singularity located at $y(F_1)$. The effect of
different (positive) values of $\Lambda$ is solely to move
the location, $y(F_1)$, of the singualarity. This is the self
tuning feature, first found in \cite{kachru}. This feature 
is thus present in the above solutions also which, however, 
differ in details from the solutions in \cite{kachru}.

{\bf 5.}  
We now conclude with a few remarks. 
In this letter, we have considered the brane world scenario in
$D = (n + m + 1)$ dimensional spacetime, with an $m$-form
antisymmetric bulk field, $m \ge 1$, and studied some of its
consequences. We found, by an analysis similar to that of
\cite{gw}, that the $m$-form field, with suitable bulk and brane
potentials, can indeed stabilise the radion modulus at the
required value with no fine tuning.  We further showed that
self tuning solutions are also present which, however, differ in
details from those in \cite{kachru}.

We are unable to solve the equations of motion analytically.
The solution generating technique of \cite{gubser}, used for the
bulk scalar, is not applicable for the bulk $m$-form field.
Clearly, it is desireable to find analytic solutions and,
thereby, to establish the stabilisation mechanism rigorously.

We assumed here that the $m$-form field has non vanishing
components along the ${\bf T^m}$ directions only. Therefore,
from the $n$ dimensional point of view, it is perhaps not
surprising that the $m$-form field mimics a scalar field. By the
same token, if the $m$-form field has non vanishing components
along $p$ of the ${\bf R^n}$ directions, $p \le Min(n,m)$, then
from the $n$ dimensional point of view, the $m$-form field will
mimic a $p$-form field. It is therefore 
important to study the phenomenological ramifications of such an
$m$-form field and, in particular, to study the signatures which 
can distinguish it from a bulk scalar field. It is also of great
interest to see if the scenario presented here can be derived 
from fundamental theories, such as 
string theory or supergravity theories. We are studying 
some of these issues at present. 




\end{document}